\documentclass[12pt,twoside]{article}
\usepackage{fleqn,espcrc1}

\newcommand{\AmS}{{\protect\the\textfont2
  A\kern-.1667em\lower.5ex\hbox{M}\kern-.125emS}}

\title{The quark distributions of baryons}

\author{Bo-Qiang Ma\address{Department of Physics,
        Peking University, Beijing 100871, China
        and Institute of High Energy Physics, Academia Sinica,
        P.~O.~Box 918(4), Beijing 100039, China}%
        \thanks{Plenary talk at
        Hadron 99, published in Nucl.~Phys.~A 675 (2000) 179-186.
        This work is partially supported by
        National Natural Science Foundation of China under Grant
        No.~19605006 and No.~19975052.}}

\begin{document}

\maketitle

\begin{abstract}
The recent progress in the investigations on the quark structure
of the $\Lambda$ and $\Sigma$ baryons are reviewed. It is shown
that the quark structure of $\Lambda$ and $\Sigma$ hyperons can
provide a new domain to test various theories concerning the
spin and flavor structure of the nucleon. The $\Lambda$ and $\Sigma$
Physics might be new directions to explore
the quark distributions of baryons both theoretically
and experimentally.
\end{abstract}

\section{Introduction}

The nucleons were considered as point-like particles when they
were first isolated in the 30's and later on they were found to have
extended structure in the 50's. The quark model
suggested the nucleons as composite systems of more basic
constituents of quarks and the lepton nucleon deep elastic scattering
experiments in the
latter 60's confirmed the quark-parton structure of nucleons and led to
a new era to consider the nucleons in terms of quark and gluon
degrees of freedom in QCD. After more than three decades experimental
and theoretical investigations of various deep inelastic scattering
processes, the structure of the nucleons
has been found to be more complicated than naively expected
and there have been many surprises found in recent years concerning the
sea quark content of the nucleons:
\begin{itemize}
\item
There have been an extensive attention on
the spin content of the proton raised by the observation of the
Ellis-Jaffe sum rule violation and a much smaller helicity
sum of quarks than naively expected \cite{SpinR,Ma98a}.
\item
The observation of the
Gottfried sum violation suggested a
flavor asymmetry of $u$ and $d$ sea quarks inside the proton
\cite{Kum97},
or alternative possibility of including also
some isospin symmetry breaking between
proton and neutron \cite{Isospin}.
\item
The current knowledge of the strange
quark content of the proton is still very poor, since
one is still unclear as to whether or not strange quarks
are highly polarized inside the proton \cite{Bro88,Bro96},
and it is even more
obscure whether or not the strange quark-antiquark distributions are
symmetric \cite{Bro96}.
\end{itemize}

It was commonly taken for granted that we understand better of the
valence quark structure of the nucleons.
However, the recent progress shows that the
flavor and spin structure of the valence quarks
is still not clear at $x=1$. For
example, there are different predictions concerning the ratio
$d(x)/u(x)$ at $x \to 1$ from the perturbative QCD (pQCD) analysis
\cite{Far75,Bro95} and the SU(6) quark-diquark model
\cite{DQM,Ma96,Mel96}, and there are different predictions
concerning the value of $F_2^n(x)/F_2^p(x)$ at large $x$, which
has been taken to be $1/4$ as in the quark-diquark model in most
parameterizations of quark distributions. Recent analysis
\cite{Mel96,Yang99} of experimental data from several processes suggests
that $F_2^n(x)/F_2^p(x) \to 3/7$ as $x \to 1$, in favor of the
pQCD prediction. The spin structure of the valence quarks is also
found to be different near $x=1$ in these models, and predictions
have been made concerning the non-dominant valence down ($d$)
quark, so that $\Delta d(x)/d(x)=-1/3$ in the quark-diquark model
\cite{Ma96,Mel96}, a result which is different from the pQCD
prediction $\Delta q(x)/q(x)=1$ for either $u$ and $d$
\cite{Bro95}. At the moment, there is still no clear data in order
to check these different predictions, although the available
measurements \cite{SMC96,HERMES99} for the polarized $d$ quark
distributions seem to be negative at large $x$, slightly in favor
of the quark-diquark model prediction.

It is important to
perform high precision measurements of available physical
quantities and/or to measure new quantities related to the flavor
and spin structure of the nucleons, in order to have a better
understanding of the quark-gluon structure of the nucleon.
However, it should be more meaningful and efficient if we can find
a new domain where the same physics concerning the structure of
the nucleons can manifest itself in a way that is more easy and
clean to be detected and checked. It has been recently suggested
by Schmidt, Yang, and I \cite{MSY2,MSY3} that
the quark structure of the $\Lambda$
and the $\Sigma$ are
such new frontiers that can be used to test various
ideas concerning the structure of the nucleons. In this talk I will
review some most recent progress in this direction and review
a case study \cite{MSY3} to check the theory on the quark distributions
of the nucleons
from the quark distributions of the $\Lambda$'s.

\section{The quark structure of the $\Lambda$}

There have been many theoretically studies concerning the
quark structure of the $\Lambda$. It was found by
Burkardt and Jaffe \cite{Bur93} that the $u$ and $d$ quarks inside
a $\Lambda$ should be negatively polarized from SU(3) symmetry. It
was also pointed out by Soffer and I \cite{Ma99} that the
flavor and spin content of the $\Lambda$ can be used to test
different predictions concerning the spin structure of the nucleon
and the quark-antiquark asymmetry of the nucleon sea.
Most
recently, Schmidt, Yang and I found \cite{MSY2,MSY3} that
the flavor and spin structure
of the $\Lambda$ near $x=1$ can provide clean tests between
perturbative QCD (pQCD) and the SU(6) quark-diquark model
predictions. We also found that the non-dominant up ($u$) and down
($d$) quarks should be positively polarized at large $x$, even
though their net spin contributions to the $\Lambda$ might be zero
or negative. There have been
calculations of the explicit shapes for the quark fragmentation
functions in a quark-diquark model \cite{Nza95}
and for the quark distributions inside the $\Lambda$
in the MIT bag model \cite{Bor99}.
The sea quark content of the $\Lambda$ has been also studied
by Boros, Londergan and Thomas and novel features in similar to the
nucleon case have been
suggested \cite{Bor99}.
Thus it is clear that the quark structure of
$\Lambda$ is a frontier which can enrich our understanding
concerning the flavor and spin structure of the nucleons and
provides a new domain to test various ideas concerning the hadron
structure that come from the available nucleon studies.

Direct measurement of the quark distributions of the
$\Lambda$ is difficulty, since the $\Lambda$ is a charge-neutral
particle which cannot be accelerated as incident beam and its
short life time makes it also difficult to be used as a target.
However, the quark distributions and the quark fragmentation
functions are interrelated quantities that can uncover the
structure of the involved hadron.
For example, we know that the quark distributions inside a hadron are
related by crossing symmetry to the fragmentation functions of the
same flavor quark to the same hadron, by a simple reciprocity
relation \cite{GLR}
\begin{equation}
q_{h}(x) \propto D_q^h(z),
\end{equation}
where $z=2 p \cdot q/Q^2$ is the momentum fraction of the produced
hadron from the quark jet in the fragmentation process, and
$x=Q^2/2 p \cdot q$ is the Bjorken scaling variable corresponding
to the momentum fraction of the quark from the hadron in the DIS
process. Although such an approximate relation may be only valid
at a specific scale $Q^2$ near $x=1$ and $z=1$ at leading order
approximation,
it can provide a reasonable connection
between different physical quantities and lead to different
predictions about the fragmentations based on our understanding of
the quark structure of a hadron \cite{Ma99,Bro97}. Thus we can use
various $\Lambda$ fragmentation processes to test different
predictions.

In principle we can test the different predictions by a
measurement of a complete set of quark to $\Lambda$ fragmentation
functions. However, in practice we do not need such systematic
studies of quark to $\Lambda$ fragmentations before we can test
the different predictions.
Many processes have been suggested to measure various quark
to $\Lambda$ fragmentation functions:
\begin{itemize}
\item
Jaffe and Burkardt \cite{Bur93} suggested one promising method to
obtain a complete set of
polarized fragmentation functions for different quark flavors
based on the measurement of the helicity asymmetry for
semi-inclusive production of $\Lambda$ hyperons in $e^+e^-$
annihilation on the $Z^0$ resonance.
\item
Measurements of the light-flavor quark fragmentations into $\Lambda$
have been also suggested from polarized electron
DIS process \cite{Jaf96} and neutrino DIS process
\cite{Kot97}, based on the $u$-quark dominance
assumption.
\item
There is also a recent interesting suggestion to determine
the polarized fragmentation functions by measuring the helicity transfer
asymmetry in the process
$p \overrightarrow{p} \to \overrightarrow{\Lambda} X$
\cite{Flo98}.
\item
More recently, Soffer and I suggested \cite{Ma99}
to measure a complete set of quark
to $\Lambda$ unpolarized and polarized fragmentation functions for
different quark flavors by the systematic exploitation of
unpolarized and polarized $\Lambda$ and $\bar{\Lambda}$
productions in neutrino, antineutrino and polarized electron DIS
processes.
\end{itemize}
Thus we have a new and rich domain from where we can study
the quark structure of the $\Lambda$ both theoretically
and experimentally.
For example, a recent detailed analysis
\cite{MSY3} of the available $\Lambda$-polarization data in
$e^+e^-$ annihilation at the $Z$-pole supports the prediction
\cite{MSY2} that
the $u$ and $d$ quarks inside the $\Lambda$ should be positively
polarized at large $x$, though their net helicities might be zero
or negative.

\section{The quark structure of the $\Sigma$}

Although the $\Lambda$ can provide a clean test of the different
flavor and spin structure between different models,
we still need a connection between the quark
distributions inside the $\Lambda$ and the quark fragmentation
into a $\Lambda$ and such a connection is not completely free from
theoretical and experimental uncertainties.
Thus it is meaningful to find
a charged baryon which has also different flavor and spin structure
between different models. The charged baryons other than
nucleons, such as $\Sigma^{\pm}$, may be used as beam to directly
measure their own quark structure in case the structure of the
target is comparatively well known.
Using the $\Sigma^{\pm}$ as beam in Drell-Yan
processes has been suggested \cite{Alberg96} for the
purpose of studying  the flavor asymmetry in the sea of the
baryons, and the sea quarks of the $\Sigma$'s have been discussed
\cite{Bor99,Alberg96}.
It has been recently found
by Schmidt, Yang and I \cite{MSY4} that the $\Sigma$'s have the most
significant difference in the flavor and spin structure of the
valence quarks between
the quark-diquark model and pQCD at medium to large $x$,
and the measurement of Drell-Yan
process for $\Sigma^{\pm}$ beams on the isoscalar targets can test
different predictions of the quark structure of the $\Sigma^{\pm}$
baryons. It is also pointed out by Cao and Signal \cite{Cao99} recently that
there are also quark-antiquark asymmetries in the $\Sigma$'s,
i.e, for $d$-$\bar{d}$
distributions inside $\Sigma^+(uus)$ and
for $u$-$\bar{u}$ distributions inside $\Sigma^-(dds)$, in analogy
to the strangeness quark-antiquark distribution asymmetry inside
the nucleon from the baryon-meson fluctuation model \cite{Bro96}.
It should be interesting if one can find a physical quantity that can
measure and test such quark-antiquark distribution asymmetries
inside the $\Sigma$'s.
Thus the quark structure of the $\Sigma$'s can be also a
new domain to test different theories concerning the quark
distributions of the nucleons.

\section{A case study}

As a case study to show that the measurement of the quark structure
of the $\Lambda$ may serve to test theory concerning the
quark structure of the nucleon, we present a brief review
on the spin and flavor structure of the valence quarks
for the nucleon and the $\Lambda$ in a light-cone
SU(6) quark-spectator-diquark model \cite{Ma96,MSY2,MSY3}.

As we know, it is proper to describe deep inelastic scattering as
the sum of incoherent scatterings of the incident lepton on the
partons in the infinite momentum frame or in the light-cone
formalism. The unpolarized valence quark distributions $u_v(x)$
and $d_v(x)$ of the nucleon are given in this model
\cite{Ma96} by
\begin{eqnarray}
&&u_{v}(x)=\frac{1}{2}a_S(x)+\frac{1}{6}a_V(x);\nonumber\\
&&d_{v}(x)=\frac{1}{3}a_V(x), \label{eq:ud}
\end{eqnarray}
where $a_D(x)$ ($D=S$ for scalar spectator or $V$ for axial vector
spectator) is normalized such that $\int_0^1 {\mathrm d} x
a_D(x)=3$, and it denotes the amplitude for quark $q$ to be
scattered while the spectator is in the diquark state $D$. Exact
SU(6) symmetry provides the relation $a_S(x)=a_V(x)$, which
implies the valence flavor symmetry $u_{v}(x)=2 d_{v}(x)$. This
gives the prediction $F^n_2(x)/F^p_2(x)\geq 2/3$ for all $x$,
which is ruled out by the experimental observation
$F^n_2(x)/F^p_2(x) <  1/2$ for $x \to 1$. The SU(6) quark-diquark
model \cite{DQM} introduces a breaking to the exact SU(6) symmetry
by the mass difference between the scalar and vector diquarks and
predicts $d(x)/u(x) \to 0$ at $x \to 1$, leading to a ratio
$F_2^n(x)/F_2^p(x) \to 1/4$, which could fit the data and has been
accepted in most parameterizations of quark distributions for the
nucleon. It has been shown that the SU(6) quark-spectator-diquark
model can reproduce the $u$ and $d$ valence quark asymmetry that
accounts for the observed ratio $F_2^{n}(x)/F_2^{p}(x)$ at large
$x$ \cite{Ma96}. This supports the quark-spectator picture of deep
inelastic scattering in which the difference between the mass of
the scalar and vector spectators is essential in order to
reproduce the explicit SU(6) symmetry breaking while the bulk
SU(6) symmetry of the quark model still holds.

The quark helicity distributions for the $u$ and $d$ quarks can be
written as \cite{Ma96}
\begin{equation}
\begin{array}{llcr}
\Delta u_{v}(x)=u_{v}^{\uparrow}(x)-u_{v}^{\downarrow}(x)
=-\frac{1}{18}a_V(x)W_q^V(x) +\frac{1}{2}a_S(x)W_q^S(x);
\\
\Delta d_{v}(x)=d_{v}^{\uparrow}(x)-d_{v}^{\downarrow}(x)
=-\frac{1}{9}a_V(x)W_q^V(x),
\end{array}
\label{eq:sfdud}
\end{equation}
in which $W_q^S(x)$ and $W_q^V(x)$ are the Melosh-Wigner
correction factors \cite{Ma91b} for the scalar and axial
vector spectator-diquark cases. From Eq.~(\ref{eq:ud}) one gets
\begin{eqnarray}
&&a_S(x)=2u_v(x)-d_v(x);\nonumber\\ &&a_V(x)=3d_v(x).
\label{eq:qVS}
\end{eqnarray}
Combining Eqs.~(\ref{eq:sfdud}) and (\ref{eq:qVS}) we have
\begin{eqnarray}
&&\Delta u_{v}(x)
    =[u_v(x)-\frac{1}{2}d_v(x)]W_q^S(x)-\frac{1}{6}d_v(x)W_q^V(x);
\nonumber    \\ &&\Delta d_{v}(x)=-\frac{1}{3}d_v(x)W_q^V(x).
\label{eq:dud}
\end{eqnarray}
Thus we arrive at simple relations \cite{Ma96} between the
polarized and unpolarized quark distributions for the valence $u$
and $d$ quarks. The calculated polarization asymmetries $A_1^N=2 x
g_1^N(x)/F_2^N(x)$, including the Melosh-Wigner rotation, have
been found \cite{Ma96} to be in reasonable agreement with the
experimental data, at least for $x \geq 0.1$. A large asymmetry
between $W_q^S(x)$ and $W_q^V(x)$ leads to a better fit to the
data than that obtained from a small asymmetry.

The key point that the light-cone SU(6) quark-diquark model
can give a good description of the experimental observation
related to the proton spin quantities relies on the fact
that the quark helicity measured in polarized deep inelastic
scattering is different from the quark spin in the rest frame
of the nucleon or in the quark model \cite{Ma98a,Ma91b}.
Thus the observed
small value of the quark helicity sum for all quarks
is not necessarily
in
contradiction with the quark model in which the proton
spin is provide by the valence quarks.
From this sense, there is no serious ``spin puzzle'' or
``spin crisis'' as it was first understood.
Of course, the sea quark
content of the nucleon is complicated and it seems that
the baryon-meson fluctuation configuration \cite{Bro96}
composes one important
part of the non-perturbative aspects of the nucleon. We should
not expect that the valence quarks provide 100\% of the proton spin,
and the sea quarks and gluons should also contribute some
part of the proton spin, thus it is meaningful to design
new experimental methods to measure these contributions independently.
Useful relations that can be used to measure the quark spin
as meant in the quark model and the quark orbital angular
momentum from a relativistic viewpoint have been discussed
in Refs.~\cite{Ma98a} and \cite{Ma98b}.
It has been pointed out \cite{Ma98a} that
the quark spin distributions
$\Delta q_{QM}(x)$ are connected
with the quark helicity distributions $\Delta q(x)$ and the
quark transversity distributions $\delta q(x)$
by an approximate relation:
\begin{equation}
\Delta q_{QM}(x) + \Delta q(x)=2 \delta q(x).
\end{equation}
The quark orbital angular momentum $L_q(x)$
and the quark helicity distribution
$\Delta q(x)$ are also found \cite{Ma98b}
to be connected to the quark model spin distribution
$\Delta q_{QM}(x)$ by
a relation:
\begin{equation}
\frac{1}{2}\Delta q(x)+ L_q(x)=\frac{1}{2}\Delta q_{QM}(x),
\end{equation}
which means that one can decompose the quark model spin
contribution $\Delta q_{QM}(x)$ by a quark helicity term
$\Delta q(x)$ {\it plus} an orbital
angular momentum term $L_q(x)$.
There is also a new relation connecting
the quark orbital angular momentum with
the measurable quark helicity
distribution and transversity distribution:
\begin{equation}
\Delta q(x)+L_q(x)=\delta q(x),
\end{equation}
from which we may have new
sum rules connecting the quark orbital angular momentum
with the nucleon axial and tensor charges.
The quark tranversity
and orbital angular momentum distributions have been also
calculated in the light-cone SU(6) quark-diquark model
\cite{Ma98a,Ma98b}. Thus future measurements of new physical
quantities related to the proton spin structure can be used
to test whether the framework is correct or not, and detailed predictions
and discussions can be found in Refs.~\cite{Ma98a,Ma98b,Ma98c}.
We point out that one of the predictions of the framework
is the small helicity contribution from the anti-quarks
and the available experimental data \cite{SMC96,HERMES99}
are consistent with this prediction. This is different from
most other works in which a large negative spin contribution
from anti-quarks is required to reproduce the observed
small quark helicity sum. In our framework the Melosh-Winger
rotation \cite{Ma98a,Ma91b} and the flavor asymmetry of the
Melosh-Wigner rotation factors between the $u$ and $d$ quarks
\cite{Ma96} are the main reason for the reduction of the
quark helicity sum compared to the naive quark model prediction.

However, as I have pointed out in the Introduction,
the quark distributions of the $\Lambda$ can be also
used as a new domain to check the light-cone SU(6)
quark-diquark model.
By applying the same analysis from the nucleon case to the
$\Lambda$ case by Schmidt, Yang, and I \cite{MSY2,MSY3},
we get the unpolarized quark distributions for the
three valence $u$, $d$, and $s$ quarks for the $\Lambda$,
\begin{equation}
\begin{array}{clcr}
&u_v(x)=d_v(x)=\frac{1}{4} a_{V(qs)}(x) +\frac{1}{12}
a_{S(qs)}(x);
\\
&s_v(x)=\frac{1}{3} a_{S(ud)}(x),
\end{array}
\end{equation}
where $a_{D(q_1 q_2)}(x) \propto  \int [\mathrm{d}^2
\vec{k}_\perp] |\varphi (x, \vec{k}_\perp)|^2$ ($D=S$ or $V$)
denotes the amplitude for the quark $q$ being scattered while the
spectator is in the diquark state $D$, and is normalized such that
$\int_0 ^1 a_{D(q_1 q_2)}(x) \mathrm{d} x =3$.
Similar to the nucleon case, the quark spin distributions for the
three valence quarks can be expressed as,
\begin{equation}
\begin{array}{clcr}
&\Delta u_v(x)=\Delta d_v(x)=-\frac{1}{12}
a_{V(qs)}(x)W_{V(qs)}(x) +\frac{1}{12} a_{S(qs)}(x)W_{S(qs)}(x);
\\
&\Delta s_v(x)=\frac{1}{3} a_{S(ud)}(x)W_{S(ud)}(x),
\end{array}
\end{equation}
where $W_D(x)$ is the correction factor due to the Melosh-Wigner
rotation. It has been shown \cite{MSY2,MSY3} that the differences in
the diquark masses $m_{S(ud)}$, $m_{S(qs)}$, and $m_{V(qs)}$ cause
the symmetry breaking between $a_{D(q_1 q_2)}(x)$ in a way that
$a_{S(ud)}(x) > a_{S(qs)}(x) > a_{V(qs)}(x)$ at large $x$.
Thus the
quark-diquark model predicts, in the limit $x \to 1$, that
$u(x)/s(x) \to 0$ for the unpolarized quark distributions, $\Delta
s(x)/s(x) \to 1$ for the dominant valence $s$ quark, and also
$\Delta u(x)/u(x) \to 1$ for the non-dominant valence $u$ and $d$
quarks.

Recently there have been detailed measurements of the $\Lambda$
polarizations from the $Z$ decays in $e^+e^-$-annihilation
\cite{ALEPH96,DELPHI95,OPAL97}. The measured
$\Lambda$-polarization has been compared with several theoretical
calculations \cite{Kot97,Bor98,Flo98b} based on simple ansatz such
as $\Delta D_{q}^{\Lambda}(z) =C_{q}(z) D_{q}^{\Lambda}(z)$ with
constant coefficients $C_{q}$, or Monte Carlo event generators
without a clear physical motivation. Schmidt, Yang and I \cite{MSY3}
calculated the $\Lambda$-polarization in
$e^+e^-$-annihilation at the $Z$-pole by connecting the
quark to $\Lambda$ fragmentation functions with the quark
distributions inside the $\Lambda$.
It has been found \cite{MSY3} that the
theoretical results from the quark-diquark model fit the data very
well for the available $\Lambda$-polarization data in
$e^+e^-$ annihilation at the $Z$-pole within its present precision.
Thus the quark-diquark model provides a successful
description of the $\Lambda$-polarization $P_{\Lambda}(z)$, in
addition to its successful descriptions of the ratio
$F^n_2(x)/F^p_2(x)$ and the polarized structure functions for the
proton and neutron.
However, pQCD can also give a good description of the
data by taking into account the suppression of quark helicities
compared to the SU(6) quark model values of quark spin
distributions. Thus the prediction of positive polarizations for
the $u$ and $d$ quarks inside the $\Lambda$ at $x \to 1$ is
supported by the available experimental data.

It is still not possible to
make a clear distinction between the two different predictions of
the flavor and spin structure of the $\Lambda$ by only the
$\Lambda$-polarization in $e^+e^-$-annihilation near the $Z$-pole.
Thus
new information from other quantities related to the flavor and
spin structure of the $\Lambda$ are needed before we can have a
clean distinction between different predictions, and it seems that
$\Lambda$ ($\bar{\Lambda}$) production in the neutrino
(anti-neutrino) DIS processes \cite{Ma99} are more sensitive to
different flavors.

\section{Conclusion}
From the above brief review on the recent progress
in the investigations on the quark structure
of the $\Lambda$ and $\Sigma$ hyperons, we arrive
at the conclusion that the quark structure
of $\Lambda$ and $\Sigma$ hyperons can
provide new domains to test various theories concerning the
spin and flavor structure of the nucleon. Thus the $\Lambda$ and $\Sigma$
Physics should be new directions to explore
the quark distributions of baryons both theoretically
and experimentally.

\vspace{5mm}
This review is based on the works with
my collaborators Stan Brodsky,
Andreas Sch\"afer, Ivan Schmidt, Jacques Soffer,
and Jian-Jun Yang. I would like to express my great
thanks to them for the
enjoyable collaborations and the encouragements from them.

\end{document}